\documentstyle[prl,aps,twocolumn,floats]{revtex}
\begin{document}
\widetext
\title{ Compressibility of a quantum Hall system at $\nu = 1/2$ }
\author{  B.~I.~Halperin$^{a}$ and  Ady Stern$^b$  }
\address{
{\it (a)} Physics Department, Harvard University, Cambridge MA 02138;
{\it (b)} Department of Condensed Matter Physics, Weizmann Institute,
Rehovot 76100, Israel}
\maketitle

\begin{abstract}
Despite the fact that composite fermions may be regarded as electrically
neutral objects at Landua-level filling factor $\nu=1/2$, carrying only an
electric dipole moment, we argue that the long-wavelength compressibility
of the system is finite.
   \end{abstract}

\bigskip
\narrowtext
Several recent articles \cite{Shankar}--\cite{Read} have emphasized
the fact that the low energy "composite fermion" (CF) excitations at
Landau-level filling fraction $\nu = 1/2$ are neutral with respect to
the electric charge, carrying only an electric dipole moment, and some
have suggested that as a consequence, the electron system at $\nu =
1/2$ should be "incompressible" in the long-wavelength limit -- i.e.,
that the zero-frequency electron density response function $\chi (q)$
should vanish $\propto q^2$, in the limit $q\rightarrow
0$.\cite{Shankar,Pasquier} We argue that despite the dipole
nature of the dressed CFs, the electron system is compressible (i.e.,
$\chi$ is finite at $q=0$ for short-range forces, and vanishes as $q/
2 \pi e^2$ for unscreened Coulomb interactions), as predicted by 
conventional fermion Chern-Simons (CS) theory \cite{HLR}--\cite{Kim}.

In Eq (17) of Ref. \cite{Pasquier}, Pasquier and Haldane find $\chi
(q) \propto m^* q^2 / (1+F_1) $ , where $m^*$ is the composite-fermion
effective mass, and $F_1$ is the first Landau interaction parameter.
We argue, however, that $F_1 \rightarrow -1$, for $q \rightarrow 0$,
in such a manner as to give a finite compressibility. In ref [1],
Murthy and Shankar suggest $\chi \propto q^2$, but also raise the
possibility that the "drifting Fermi sea", peculiar to this system
"might lead to soft modes and compensating inverse powers of $q$."  We
argue that the latter does occur. The finite compressibility is
restored because of the peculiarity of the CF system, noted previously
by Haldane, and implicit in \cite{Shankar}, that {\it the energy of
  the system is unchanged if a constant ${\bf K}$ is added to the momentum of
  every CF}.  Within the formalism of \cite{Pasquier}, the fact that
$(1+F_1)=0$ , for $q=0$, follows directly from this constraint and the
definition of the Landau parameters.

We employ here the formalism of \cite{Shankar}, where a unitary
transformation is used to approximately decouple the fermion degrees of
freedom from bosonic oscillators, describing cyclotron motion. The
${\bf K}$-invariance of the energy follows there from gauge invariance.  The
fermion part of the approximate Hamiltonian (Eq. (27) of [1]) is, in
the absence of electron-electron interactions,
\begin{equation}
H = \sum_i {{\bf p}^2_i \over 2m} -  \sum^{\bf Q}_{{\bf k}=0} \:  \sum_{i,j}
{{\bf p}_i  \cdot {\bf p}_j \over 2mn } \: e^{-i {\bf k} \cdot 
({\bf r}_i -{\bf r}_j)}
\label{ham}
\end{equation}
where $m$ is the electron mass, and $n$ is the average density. The
required invariance is satisfied if one keeps only the $k=0$ term in
the second sum; fluctuation contributions from $k \not= 0$ should be
cancelled by higher order terms which were omitted from (\ref{ham}).
In Ref. \cite{Shankar}, the magnitude of the upper cutoff $ Q$ was
chosen equal to $k_F$, in which case the inverse effective mass
$1/m^*$, obtained by adding the $i=j$ terms of the second sum to the
bare kinetic energy, is equal to zero -- a correct feature of the
non-interacting electron system. The price paid for that is the
violation of the ${\bf K}$--invariance of the energy by the $k\ne 0$
terms in (\ref{ham}).  Here, we consider a model in which Q is much
less than $k_F$, such that there are only a few $k\ne 0$ terms.  This
model corresponds to a fermion-CS model where the CS charges are
smeared over a radius of order $Q^{-1}$. \cite{HH} In this model, the
effective mass $m^*$, obtained by the method of [1], is finite, even
in the absence of electron--electron interaction; however, the long
wavelength features that we are interested in are unaffected.  In
particular, the transformed composite fermions are still electrically
neutral objects, and the system is still invariant under addition of a
constant ${\bf K}$ to the momentum of every fermion.

We now show that the $K$--invariance of the fermion energy leads to
a finite compressibility of the system.  Since a CF of momentum
${\bf p}$ carries a dipole moment $l_0^2\hat z \times {\bf p}$ (where
$l_0$ is the magnetic length), the CF contribution to the
electron-density operator
is given by $ - il_0^2 {\bf q} \times {\bf g}({\bf q})$,
with ${\bf g}({\bf q}) = \sum {\bf p}_j e^{-i {\bf q} \cdot {\bf r}_j}$
being the momentum density of the CFs.  
Consequently, for ${\bf q}||{\hat x}$ the
response function $\chi$ for the electron density is propotional to
$q^2 \Phi_{yy}(q)$, where $\Phi_{\mu \nu}(q)$ describes the response
of the cartesian component $g_{\mu}$ to a perturbation of form $-
\lambda g_{\nu}(q)$.

The response function $\Phi_{yy}(q)$ is related to the energy cost
involved in a static fluctuation $g_{\mu}(q)$, for $\lambda=0$. This
cost is $ \propto |g_{\mu}|^2 \Phi^{-1}_{\mu \mu}(q) $.  For a normal
fermi liquid, gauge invariance dictates that $\Phi_{xx} = nm \equiv
\Phi_0 $, independent of the magnitude of $q$.  The transverse part,
however, satisfies $\Phi^{-1}_{yy} = \Phi^{-1}_0 + Dq^2$, where $D$ is
proportional to the diamagnetic susceptibility. For the CF system, as
stated above, it costs no energy to introduce a long wave--length
longitudinal fluctuation in ${\bf g}$, so $\Phi_0^{-1} = 0$.  If we
assume that $\Phi^{-1}$ is regular, for $q \rightarrow 0$, we find
that for the transverse component, $\Phi^{-1}_{yy}(q) \propto q^2$, so
$\chi$ is finite, and the $\nu=1/2$ state is compressible.

The above scenario has been verified by more detailed calculations, 
within the formalism of Ref. \cite{Shankar}, in the limit  $q < Q \ll k_F$,
for non-interacting particles and for the 
case of the case of weak short-range repulsion.
For a $1/r$ Coulombic repulsion, one finds $\chi \propto q$ as
expected \cite{gof}. 


Acknowledgments: We are indebted to F. von Oppen, R. Shankar and S. Simon for
instructive discussions. Work was supported by NSF grant DMR94-16910,
Israel-US Binational Science Foundation grant (95/250-1) and a Minerva
foundation grant.


\begin{thebibliography}{10}

\bibitem{Shankar}
 R. Shankar and G. Murthy, Phys. Rev. Lett. 79, 4437 (1997), and
unpublished work.

\bibitem{Pasquier} V. Pasquier and F. D. M. Haldane, cond-mat/9712169

\bibitem{DHLee} D.-H. Lee, cond-mat/9709233

\bibitem{Read} N. Read, Semicond. Sci. Tech. 9, 1859 (1994) 

\bibitem{HLR} B. I. Halperin,  N. Read, and P. A. Lee, Phys. Rev. B47, 7312 (1993 
).

\bibitem{Kim} Y. B. Kim et al, Phys. Rev. B52, 17275 (1995); A. Stern and B. I. Halperin,
Phys. Rev. B52, 5890 (1995)

\bibitem{HH} B. I Halperin, Phys. Rev. B 45, 5504 (1992); A. R. Chari, F. D. M.
Haldane, and K. Yang, cond-mat/9707055
\bibitem{gof} Manuscript in preparation
\end{thebibliography}
\end{document}